# The thermal stability and separation characteristic of anti-sticking layers of Pt/Cr films for hot slumping technology[*]

MA Shuang(马爽), WEN Ming-Wu(闻铭武), WANG Zhan-Shan(王占山)[**]

MOE Key Laboratory of Advanced Micro-Structured Materials, Institute of Precision Optical Engineering, School of Physics Science and Engineering, Tongji University, Shanghai 200092, China

The thermal stability and separation characteristic of anti-sticking layers of Pt/Cr films were studied in this paper. Several types of adhesion layers were investigated: 10.0 nm Pt, 1.5 nm Cr+50.0 nm Pt, 2.5 nm Cr+50.0 nm Pt and 3.5 nm Cr+50.0 nm Pt fabricated using direct current magnetron sputtering. The variation of layer thicknesses, roughness, crystallization and surface topography of Pt/Cr films have been analyzed by grazing incidence X-ray reflectometry, large angle X-ray diffraction and the optical profiler before and after heating. 2.5 nm Cr+50.0 nm Pt films exhibit the best thermal stability and separation characteristic according to the heating and hot slumping experiments. The films were also applied as anti-sticking layers to optimize the maximum temperature of hot slumping technology.

**hot Slumping technology, anti-sticking layers, Pt/Cr film, thermal stability, separation characteristic**

**PACS: 95.55.Ka**

## 1 Introduction

Wolter-I type grazing incidence X-ray telescopes can be used to produce images of X-ray radiation from supernova remnants, clusters of galaxies, active galactic nuclei and other objects [1-9]. Substrate fabrication and system alignment are two main factors that limit the resolution of grazing incidence X-ray telescopes [10]. There are four kinds of methods to fabricate the substrate of telescope: super polishing, nickel electroplating, epoxy replicated and hot slumping technology [11-13]. Hot slumping is an advanced technology currently with high efficiency, good stability and reproducibility [14-20], which allows shaping a sheet of glass on the mould of desired shape by applying a suitable thermal cycle.

During the progress of thermal cycle, the glasses are typically heated to several hundred degrees Celsius and then the glasses slumps over the mould with the gravity. To overcome the sticking problem between glass and mould, a suitable anti-sticking layer is necessary. American group coated BN as anti-sticking layers on the mould, which exhibits good anti-adhesion properties [21]. However, using the material of BN might increase mid-frequency errors due to its dusty consistency [22]. As a comparison, Pt has a series of advantages such as high density(21.45 g·cm$^{-3}$), high melting point(1773℃) and good chemical stability, however, a vital problem that must be solved is the relatively low adhesion of Pt layer on the Fused Silica mould. In particular, a thin Cr interlayer may improve the reliability of Pt layer. Italy group has been successfully added 5 nm Cr layer between 50 nm Pt layer and the mould in the hot slumping technology [19]. However, we find that there are some defects in surface of glass mirror using the anti-sticking layers of 5 nm Cr +50 nm Pt. The reasons why they used anti-sticking layers of 5 nm Cr +50 nm Pt are not given and the detailed layer structure and stability properties when using different thickness of Cr and Pt have not been studied yet.

In this paper, we will further research quantitative experiment and the anti-adhesion properties of Pt/Cr layer. The purpose of this paper was to determine a suitable anti-sticking layer and fibrate the substrate for Wolter-Ⅰ X-ray telescope developed for the Chinese Academy of Sciences (CAS) [23, 24]. The hot slumping technology was used for substrate fabrication. This paper mainly analyses thermal stability and separation characteristic of these anti-sticking layers by the heating experiments and the metrology results.

---

[*] supported by CAS XTP project XDA04060605

[**] corresponding author, E-mail: wangzs@tongji.edu.cn

## 2  Experiment details

There are two kinds of mould: plane and spherical mirrors. Different anti-sticking layers were deposited using direct current magnetron sputtering. The film structures consist of 10.0 nm Pt, 1.5 nm Cr+50.0 nm Pt, 2.50 nm Cr+50.0 nm Pt, and 3.5 nm Cr+50.0 nm Pt. These deposited plane and spherical mirrors were used as moulds in the process of the hot slumping.

After deposition of the anti-sticking layers, we conducted two experiments: hot slumping and heating experiment, which have the same experimental parameters and processes with or without glass samples on the surfaces of the moulds. The separation characteristic of these anti-sticking layers were studied via hot slumping experiment which placed a flat glass sheet on a coated spherical mandrel in order to obtain the desired shape by a thermal treatment inside an oven. The experiment temperature was ramped up from room temperature to about 557℃ and maintained for two hours, then the mirrors were cooled slowly to room temperature. Heating experiment would be used to study the thermal stability of these layers.

The anti-sticking layers of Pt and Pt/Cr films were first characterized by grazing incidence X-ray reflectometry (GIXR) using a lab-based diffractometer (D1 system, Bede Inc.) with the Cu-K$\alpha$ line as the source ($\lambda$ = 0.154 nm) before and after thermal treatment. The variation of film thickness and interface roughness can be determined from fits to the reflective curves. Large angle X-ray Diffraction (XRD) analyses were performed on the other diffractometer (D8 Advance, Bruker) to obtain the crystalline structure of the coatings, which can be calculated from peak 2Theta-position value and full width at half maximum (FWHM) of the peak. We also use the optical profiler (Contour GT-X3 from Bruker) to describe the surface roughness of films covering the measurement area of 62 μm×47 μm.

## 3  Results and discussion

### 3.1  The adhesive capacity tests

A direct hot slumping technology, in which the glass sheet slumps over the desired coated mould with the only action of gravity in a maximum temperature of 557 ℃, was employed to verify the separation characteristics of all the adhension layers. Fig. 1 give the photographs of the coated moulds and glasses after the hot slumping. For the 10.0 nm Pt and 1.5 nm Cr+50.0 nm Pt, parts of Pt films migrated to the glass surfaces, which indicates the adhesive force between the moulds and the films is not large enough When increasing the thickness of Cr layer to 2.5 nm and 3.5nm, the adhesive force between films and the moulds increased significantly. The moulds with 2.5 nm Cr+50.0 nm Pt and 3.5 nm Cr+50.0 nm Pt films were successfully seperated from the glasses.

In order to study the thermal stability of these layers, heating experiment and measurement instruments would be used to research the variation of film thickness, interface roughness, crystalline structure and surface roughness of these layers that coated in plane silicon mirrors.

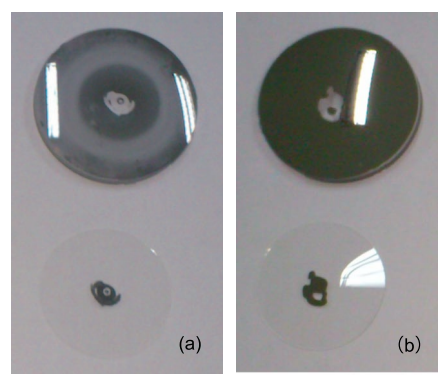

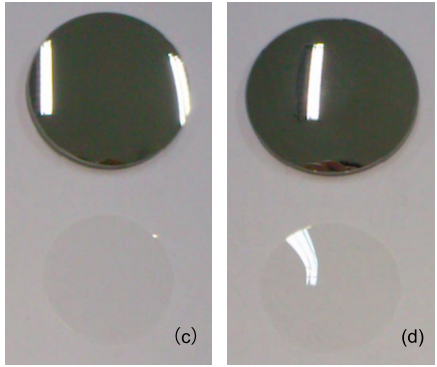

**Figure 1.** The photographs of mould-glass sticking tests after hot slumping：(a)10.0 nm Pt；(b)1.50 nm Cr+50.0 nm Pt (c) 2.50 nm Cr+50.0 nm Pt (d) 3.50 nm Cr+50.0 nm Pt

## 3.2 Grazing Angel X-ray Reflectometry analysis

In Fig.2, the black and blue solid lines are the measurement results of GIXRR (Grazing Incidence X-ray Reflectivity) before and after heating respectively, and the red and green solid lines are the corresponding fitting results. As shown in Fig. 2(a), the thickness of Pt single layer is 10.0 nm, the Bragg peaks positions almost remained unchanged, while the intensity decreased after heating, which indicate that the interface roughness of the 10.0 nm Pt layer increased during the heating. As for 1.5 nm Cr+50.0 nm Pt in Fig. 2(b), both the intensity and peaks positions are approximately the same, the results would illustrate that the film structure has little change, thus this film has good thermal stability. When the thickness of Cr increased to 2.5 nm, the height of the diffraction peaks decreased, indicating that the interface roughness of the film increased after heating. As the thickness of film increase to 3.5 nm, there are no obvious peaks in Fig. 2(d), which shows the structure of the film has been destroyed. On the other hands, the critical angle of the total reflection becomes smaller, which indicates that the density of the film decreases after heating. The interdiffusion between Pt layer and Cr layer increases. As a result, the intensity of all the Bragg peaks decrease and the oscillation of the peaks gradually disappear.

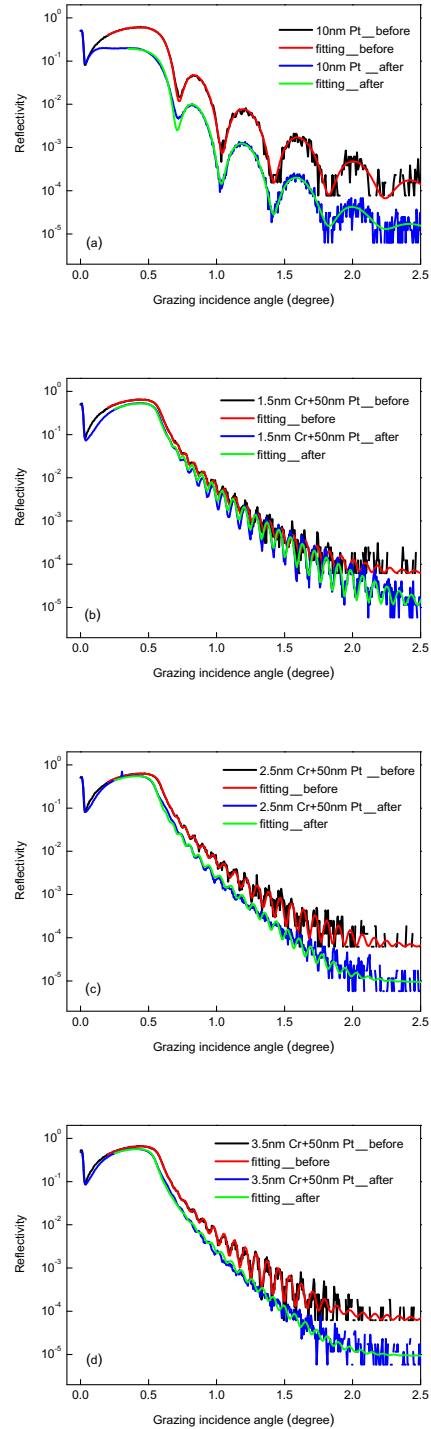

**Figure 2.** The curves of GIXRR before and after heating: (a)10.0 nm Pt；(b)1.50 nm Cr+50.0 nm Pt (c) 2.50 nm Cr+50.0 nm Pt (d) 3.50 nm Cr+50.0 nm Pt

Fitting results of film thicknesses and interface roughness are given in Table 1. Before and after heating, each thickness of the films is almost a constant. The interface roughness of 10.0 nm Pt layer has significantly increased after heating, however the interface roughness of 1.5 nm Cr+50.0 nm Pt layer is almost the same. As the thickness of Cr layer increases, the interface roughness of film also increased.

**Table 1.** Film thickness and interface roughness before and after heating

| Film structures | Pt - layer thickness(nm) | | Cr- layer thickness(nm) | | Pt- interface roughness (nm) | |
|---|---|---|---|---|---|---|
| | Before | After | Before | After | Before | After |
| 10.0 nm Pt | 10.20 | 10.16 | | | 0.38 | 0.53 |
| 1.5 nm Cr+50.0 nm Pt | 47.18 | 47.57 | 1.68 | 1.70 | 0.60 | 0.60 |
| 2.5 nm Cr+50.0 nm Pt | 47.69 | 48.33 | 2.38 | 2.46 | 0.61 | 0.73 |
| 3.5 nm Cr+50.0 nm Pt | 47.82 | 48.61 | 3.49 | 3.59 | 0.62 | 0.76 |

### 3.2 The optical profiler analysis

The optical profiler is used to describe the surface roughness of films covering the measurement area of 62 μm×47 μm. The results of surface roughness are showed in Table 2. The surface roughnesses of 10.0 nm Pt and 3.5 nm Cr+50.0 nm Pt have almost doubled after heating respectively, while the changes of surface roughnesses of 1.5 nm Cr+50.0 nm Pt and 2.5 nm Cr+50.0 nm Pt were not obvious.

**Table 2.** The surface roughness before and after heating

| Roughness $R_q$(nm) | Before heating | After heating |
|---|---|---|
| 10.0 nm Pt | 0.43 | 1.09 |
| 1.5 nm Cr+50.0 nm Pt | 0.61 | 0.72 |
| 2.5 nm Cr+50.0 nm Pt | 0.69 | 0.77 |
| 3.5 nm Cr+50.0 nm Pt | 0.57 | 0.95 |

### 3.3 Large Angle X-ray Diffraction analysis

XRD analyses were used to investigate the crystallization of the films before and after heating. Fig.3 shows the XRD patterns that were collected in the 2Theta ranges of 25°-60°. Weak reflections from Si(201) and Si(111) are observed in the part of XRD patterns. Pt films appeared strongly peaks oriented along (111) plane at all times, and absence of compound of Pt and Cr. For the thin metal films, (111) orientations from much more readily than other orientations because (111) orientations minimize the surface energy.

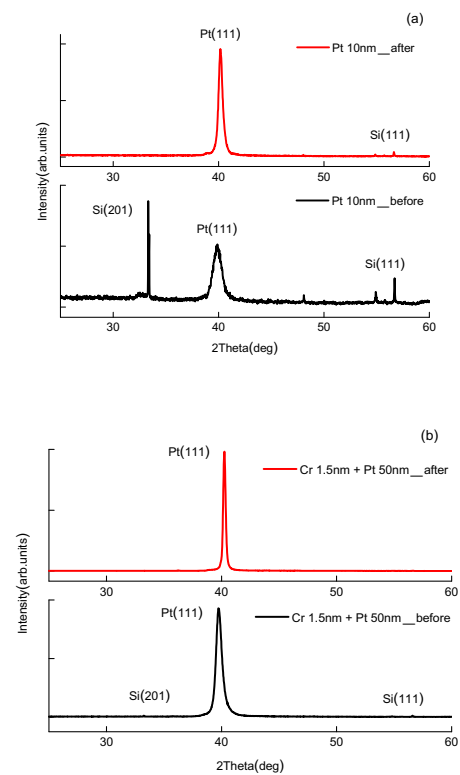

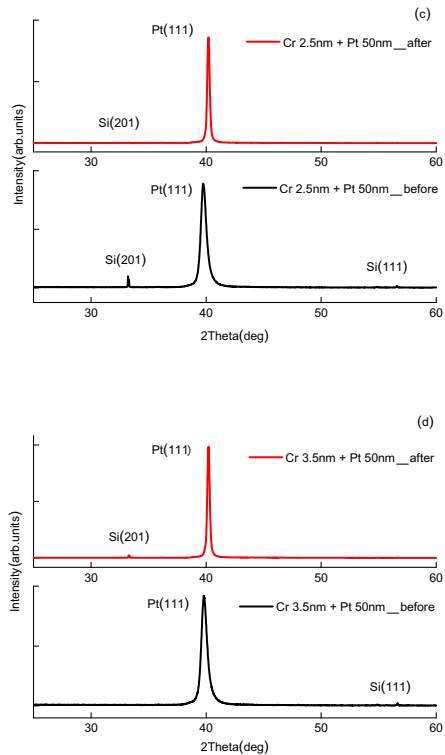

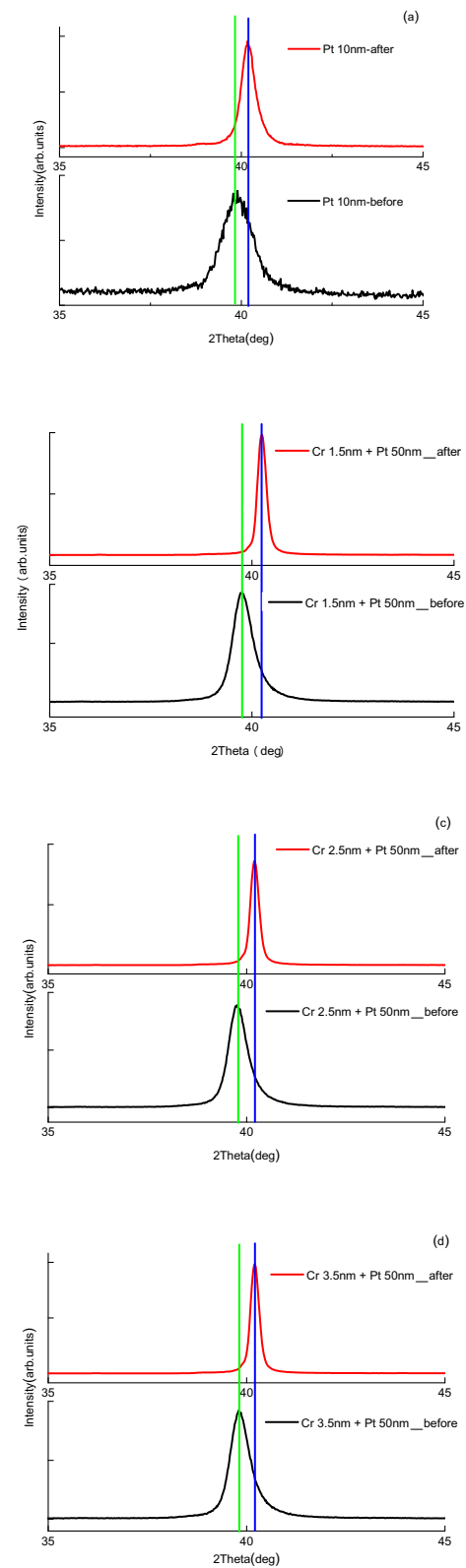

**Figure 3.** The XRD patterns (2θ: 25°-60°) before and after heating: (a)10.0 nm Pt; (b)1.50 nm Cr+50.0 nm Pt (c) 2.50 nm Cr+50.0 nm Pt (d) 3.50 nm Cr+50.0 nm Pt

Fig. 4 shows the step-scan XRD patterns (2θ: 35°-45°) of Pt(111) lines, all the Pt(111) peaks become shaper and the peak positions shift toward higher 2Theta after heating. Table 3 exhibits the fitted results of the FWHW, average crystallite size and the lattice constant d. The results show that the Pt(111) peaks of single Pt layers have broader FWHM and lower crystalline than peaks of Cr+Pt layers. After heating, the Pt(111) peaks were shifted to approximately 0.4° to larger angle, and the FWHM was reduced by half, the lattice constant decreased slightly and the average grain size increased obviously. The lattice constants of Pt(111) before heating were found to be slightly larger than the bulk Pt(d=0.2265 nm), which indicates that these films have residual compressive strains before heating. The lattice constant decreased, thus the residual compressive strain of all the films transfer to tensile strain after heating.

**Figure 4.** The XRD patterns (2θ: 35°-45°) before and after heating: (a)10.0 nm Pt; (b)1.50 nm Cr+50.0 nm Pt (c) 2.50 nm Cr+50.0 nm Pt (d) 3.50 nm Cr+50.0 nm Pt

Table 3. Correlation parameters of Pt(111) via XRD analyses

| Pt(111) | 2Theta(°) | | FWHM(°) | | d(nm) | | Size(nm) | |
| --- | --- | --- | --- | --- | --- | --- | --- | --- |
| | Before | After | Before | After | Before | After | Before | After |
| 10.0 nm Pt | 39.655 | 39.970 | 1.021 | 0.468 | 0.2271 | 0.2254 | 8.300 | 18.500 |
| 1.5 nmCr+50.0 nmPt | 39.554 | 40.009 | 0.720 | 0.233 | 0.2277 | 0.2252 | 11.800 | 40.100 |
| 2.5 nmCr+50.0 nmPt | 39.628 | 39.967 | 0.613 | 0.243 | 0.2272 | 0.2254 | 14.000 | 38.200 |
| 3.5 nmCr+50.0 nmPt | 39.673 | 39.977 | 0.625 | 0.239 | 0.2270 | 0.2253 | 13.700 | 38.900 |

### 3.4 Thermal stability analysis

According to the previous analyses, the average crystallite size of 10.0 nm Pt films increased by 10.20 nm, while the interface roughness was 0.15 nm larger and the surface roughness was increased of 0.66 nm. These are due to the destruction of the layer surfaces. When the temperature increases to a certain value, the velocity of Pt atoms also increases with the increasing temperature, which may lead to an increasing roughness of Pt film.

The average crystallite size is larger with increasing thickness of Pt layer. The thermal expansion coefficient of Cr is lower than Pt, and Cr layer hinders the diffusion of Pt layer, which explains why the average crystallite sizes of 1.5 nm Cr+50.0 nm Pt increased and the roughness changes little. This indicates that the 1.5 nm Cr + 50.0 nm Pt has good thermal stability, but it couldn't complete separation the mould and glass mirror during the hot slumping experiment. The average crystallite sizes and the interface roughness of 2.5 nm Cr+50.0 nm Pt film were increased of 24.2 nm and 0.12 nm respectively. However, the surface roughness of the film almost remains unchanged, which shows this film also has good thermal stability. As the thickness of Cr layer increased to 3.5 nm, the mixing between crystalline state of Pt and amorphous state of Cr became increases, resulting in the increase of roughness, and the interface becomes fuzzy. The layer structure was destroyed according to the GIXRR analyses, which indicates a poor thermal stability of 3.5 nm Cr+50.0 nm Pt film.

According to our analyses, 2.50 nm Cr+50.00 nm Pt film exhibits good ability of seperation and thermal stability at the same time, thus the following hot slumping experiments will be carried out with the film of 2.5 nm Cr+50.0 nm Pt as the anti-sticking layers.

### 4. Analysis the maximum temperature of hot slumping experiment

In section 3, 2.5 nm Cr+50.0 nm Pt film was used as the anti-sticking layers to successfully separate the mould and the glass. In the process of hot slumping, a proper temperature curve needs to be selected for the hot slumping thermal cycle [25]. The maximum temperature is one of the crucial parameters that have to be optimized to obtain a figured glass with low mid-spatial-frequency (MSF) error and high stability.

In this section, 2.5 nm Cr+50.0 nm Pt anti-sticking layers were used to improve the hot slumping technology. There are many visible defects on the surface of glass mirror when the temperature above 560 ℃, and the bending degree is not enough when the temperature below 550 ℃. The maximum temperatures of 557 ℃, 553 ℃ and 550 ℃ were used respectively. The surface morphology of flat glass (before heating treatment, room temperature) and figured glass (after heating treatment) was characterized using an optical profiler (Contour GT-X3 from Bruker). The PSD curves reflect the proportions of various spatial frequency components. The main coverage frequency range of the 2.5× objective lens of the optical profiler is MSF, while 10× and 50× correspond to high spatial frequency (HSF), Fig.7 shows the results of PSD curves by the objective lens: 2.5×, 10× and 50× respectively.

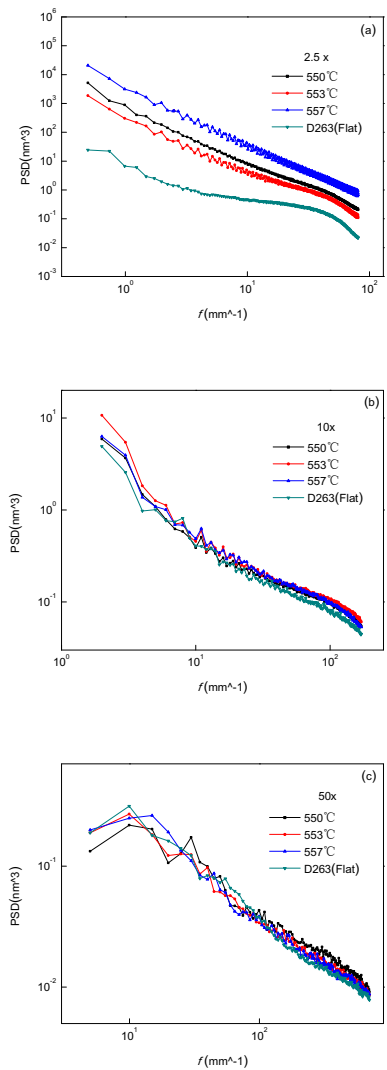

**Figure 5.** PSD curves of the figured and flat D263 glasses by different objective lens: (a) 2.5×, (b) 10×, (c) 50×.

As shown in Fig.5 (b) and (c), the PSD curves of different heating temperature are similar before and after heating, which indicates that the heating process does not affect the HSF surface roughness. In Fig.5(a), the PSD values in the temperatures of 557℃, 553℃ and 550℃ are larger than the values of room temperature(D263 Flat), which indicates that the MSF errors significantly increased after heating. The deformation of the glass surface is obviously increased when the maximum temperature is set to 557℃. The values of PSD curves between 553℃ and 550℃ are very closed. However, the smoothness of PSD curves in 550℃ is similar to the results in room temperature. In consequence, the best maximum temperature is considered to be 550℃, we will use this temperature as the maximum one in the future experiment.

## 5. Conclutions

We have investigated the characters of Pt and Pt/Cr films that acted as release agents between glasses and silicon moulds during the process of the hot slumping technology. The lattice constants of Pt(111) before heating were found to be slightly larger than the bulk Pt (d=0.2265 nm) and the films were in a state of residual compressive strains. The process of heating improved crystallite sizes of the films while decreased the lattice constants. Meanwhile a tensile strain generated inside the films. There is no direct correlation between films roughness and crystallite sizes according to our experiments. Through the hot slumping experiments, 2.5 nm Cr+50.0 nm Pt films exhibited favorable thermal stability and separation characteristic, which were used as the anti-sticking layers to improve the hot slumping technology. The best maximum temperature was optimized to be 550℃.